\documentclass[journal,twoside,final]{IEEEtran}
\normalsize
\usepackage[draft]{graphicx}
\usepackage{balance}
\usepackage{amsthm}
\usepackage{amsmath,esint}
\usepackage[mathscr]{euscript}
\usepackage{bigints}
\usepackage[FIGTOPCAP]{subfigure}
\usepackage{cite}
\usepackage{xcolor}
\usepackage{amssymb}
\usepackage{multirow}
\usepackage{nicematrix}
\usepackage{mathrsfs}
\usepackage{amsfonts}
\usepackage{psfrag}
\usepackage{tipa}
\usepackage{array}
\usepackage{soul}
\usepackage{setspace}
\usepackage{textgreek}
\usepackage{bm}
\usepackage{hyperref}
\usepackage[pdf]{epsfig}

\theoremstyle{plain}
\newtheorem{theorem}{Theorem}
\newtheorem{corollary}[]{Corollary}

\begin{document}

\title{Internodal Distance Distributions  for Static and Mobile Nodes in 2D/3D Wireless Networks}
\author{Nicholas Vaiopoulos,~Alexander~Vavoulas,~Harilaos~G.~Sandalidis,~Konstantinos~K.~Delibasis, ~and
~Dimitris~Varoutas 
\IEEEcompsocitemizethanks{\IEEEcompsocthanksitem Nicholas Vaiopoulos,~Alexander~Vavoulas,~Harilaos~G.~Sandalidis,~and~Konstantinos~K.~Delibasis are with the Department of Computer Science and Biomedical
Informatics, University of Thessaly, Papasiopoulou 2-4, 35131, Lamia, Greece. e-mail:\{nvaio,vavoulas,sandalidis,kdelimpasis\}@dib.uth.gr.}
\IEEEcompsocitemizethanks{Dimitris Varoutas is with the Department of Informatics
and Telecommunications, University of Athens, Panepistimiopolis,
15784, Athens, Greece. e-mail: arkas@di.uoa.gr.}

}
\maketitle

\begin{abstract} 
This letter presents a unified analytical framework for deriving exact closed-form internodal distance distributions in both two- and three-dimensional wireless networks, covering all four node-pair combinations (static--static, static--mobile, mobile--static, and mobile--mobile) under uniform and random waypoint spatial models. Unlike existing works that typically address isolated scenarios or rely on numerical integration and simulations, the proposed approach provides a single tractable formulation that accommodates both unequal and equal region radii and captures mobility-induced spatial effects. 
\end{abstract}


\section{Introduction}

\IEEEPARstart{I}{nternodal} distance distributions are particularly critical in random networks, where node positions are determined by stochastic models. In these systems, randomness often stems from unpredictable node placement, user mobility, and environmental factors \cite{J:Moltchanov}. Their accurate knowledge is essential for the effective analysis and design of wireless communication protocols, including routing, resource allocation, power control, and capacity planning \cite{B:Haenggi}.

To capture realistic operational constraints, it is often assumed that wireless nodes operate within bounded spatial regions. Circular and spherical domains in two- and three-dimensions (2D/3D) provide natural models for such regions when no direction is privileged, as in cellular coverage areas, indoor environments, aerial base stations, and clustered sensor deployments around a central gateway \cite{J:Badarneh,J:Gupta}. In addition, many practical systems involve multiple classes of nodes organized around a common reference point but operating over different spatial extents. This leads to concentric regions with unequal radii, which arise in relay- or unmanned aerial vehicles (UAV)-assisted communications, device-to-device links within a cell, and underwater networks \cite{J:Pan}, \cite{J:Chetlur}.

Several previous studies have derived distance distributions for nodes in bounded regions; however, a unified analytical framework is still lacking. Nichols \textit{et al.} in \cite{J:Nichols}, analyzed uniform 3D deployments, while mobility effects have been explored using random waypoint (RWP) \cite{J:Hyytia} and random walk (RW) models. Zhong \textit{et al.} \cite{J:Zhong} derived the integral-form probability density functions (PDFs) and the cumulative distribution functions (CDFs) of distances for static–mobile node pairs in 2D domains, but extensions to other scenarios and 3D geometries remain limited. 

This letter presents a unified framework for exact internodal distance distributions in 2D and 3D wireless networks under uniform and RWP models. The key assumptions underlying the analysis are detailed in the next section.

\section{Model Assumptions}
We assume that node locations are statistically independent and confined within concentric circular (2D) or spherical (3D) regions of radii $R_1$ and $R_2>R_1$. Static nodes are uniformly distributed within their respective regions, while mobile nodes\footnote{In this work, the term \textit{mobile node} refers to a node whose location follows the stationary spatial distribution induced by the RWP, rather than its explicit time evolution or trajectory.} follow an RWP model with zero pause time and are characterized by their stationary spatial distribution\footnote{The RWP model is utilized only through its stationary spatial distribution, ignoring temporal correlation.}.

Four representative user distribution scenarios are defined for both 2D and 3D environments:
\begin{itemize}
  \item \textbf{Scenario 1 (s1):} The mobile node lies within a circular (or spherical) region of radius $R_1$ and the static node within a radius $R_2$.
  \item \textbf{Scenario 2 (s2):} The static node is confined to a radius $R_1$ and the mobile node to a radius $R_2$.
  \item \textbf{Scenario 3 (s3):} Both nodes are mobile, each confined within regions of radii $R_1$ and $R_2$, respectively.
  \item \textbf{Scenario 4 (s4):} Both nodes are static within regions of radii $R_1$ and $R_2$, respectively, as examined in~\cite{B:Mathai} for the 2D case only.
\end{itemize}

 The asymmetric case $R_1 \neq R_2$ models heterogeneous deployments with different node types operating over different spatial extents, e.g., indoor users and UAV relays.  A special case with $R_1 = R_2$ is also considered, as studied in several scenarios such as \cite{J:Zhong,B:Mathai, J:Bettstetter, J:Parry}. Under these assumptions, the analytical expression for the PDF of the internodal distance, $r$, is obtained as

\begin{equation}
  f_{\mathbf{r}}(r)=\int_{0}^{\infty}  f_{\mathbf{r}/\bm{\rho}}(r/\rho)f_{\bm{\rho}}(\rho)d\rho,
  \label{CondProb} 
\end{equation}
where $f_{\mathbf{r}/\bm{\rho}}(r/\rho)$ denotes the conditional PDF of $r$, given that node 1 is located at a distance $\rho$ from the center. The term $f_{\bm{\rho}}(\rho)$ represents the PDF of the distance of node 1, corresponding either to the RWP model or a uniform distribution, depending on the scenario under consideration.

\begin{table*}[t]
\centering
\caption{Definition of $r$ and $\rho$ Intervals for the 2D and 3D cases.}
\label{Table1} 
\resizebox{\textwidth}{!}{%
\renewcommand{\arraystretch}{1.5}
\begin{NiceTabular}{c c c c c}[hvlines]  
    \hline \hline
    \large \textbf{Interval I}: $0<r<R_2-R_1$ 
    & \multicolumn{2}{c} {\large \textbf{Interval II}: $R_2-R_1<r<R_{2}$}  
    & \multicolumn{2}{c} {\large \textbf{Interval III}: $R_{2}<r<R_1+R_2$}  \\ 

    \large $0<\rho<R_{1}$ 
     & \large $0<\rho<R_{2}-r$  & \large $R_{2}-r<\rho<R_{1}$  
    & \large $0<\rho<r-R_{2}$  & \large $r-R_{2}<\rho<R_{1}$ \\   

    \includegraphics[keepaspectratio,width=1.5in]{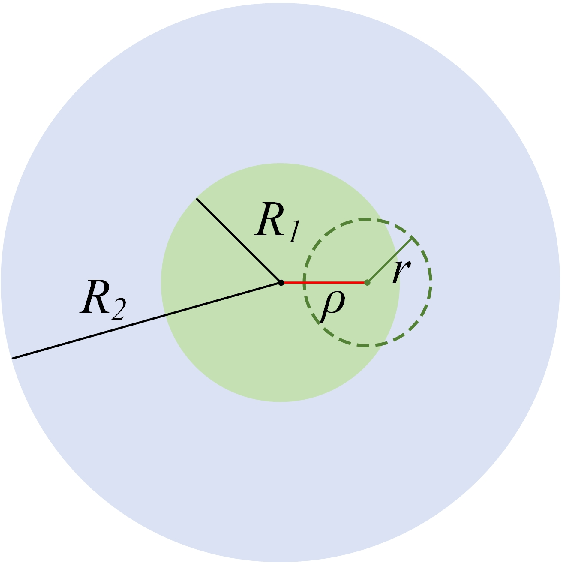}  
    & \includegraphics[keepaspectratio,width=1.8in]{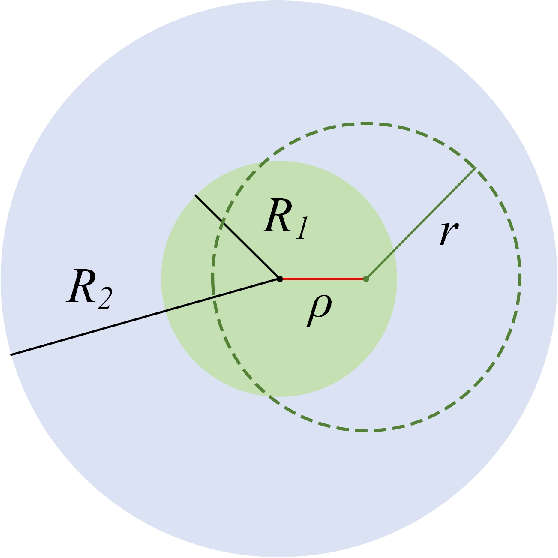}   
    &  \includegraphics[keepaspectratio,width=1.8in]{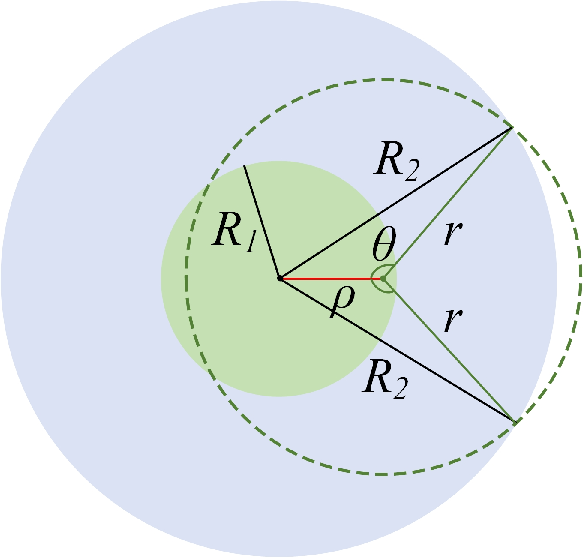}  
    & \includegraphics[keepaspectratio,width=1.8in]{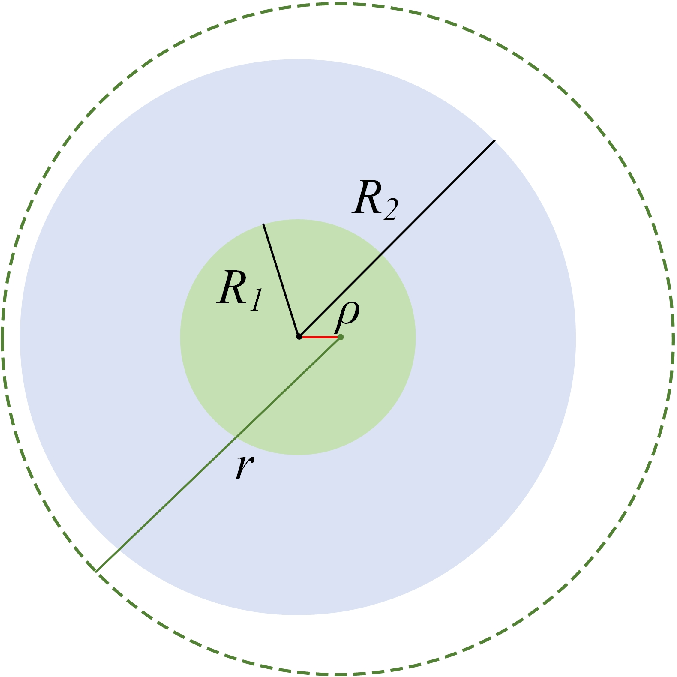}   
    &  \includegraphics[keepaspectratio,width=1.8in]{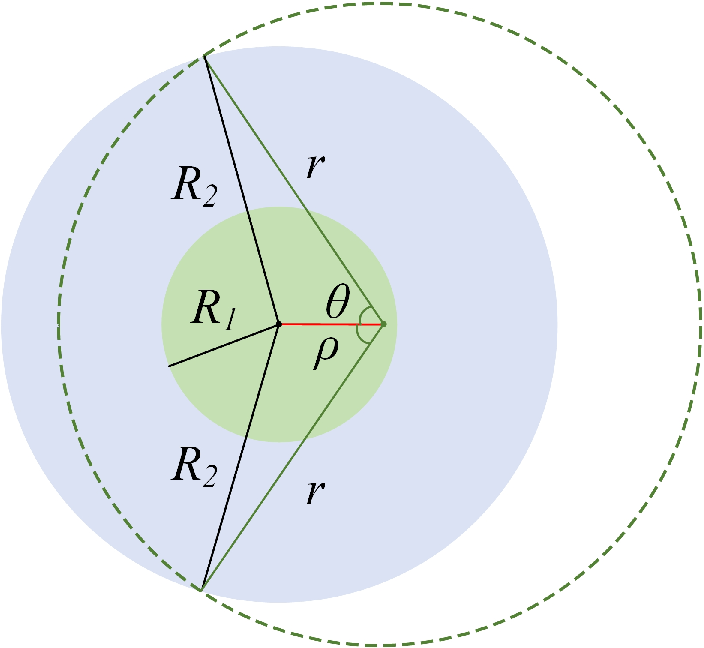} \\ 

   \large $f_{r}(r)=\int_{0}^{R_{1}} g_{1}(\rho,r)f_{\rho}(\rho) \,d\rho$ 
    & \multicolumn{2}{c} {\large $f_{r}(r)=\int_{0}^{R_{2}-r} g_{1}(\rho,r)f_{\rho}(\rho) \,d\rho + 
        \int_{R_{2}-r}^{R_{1}} g_{2}(\rho,r)f_{\rho}(\rho) \,d\rho$}  
    & \multicolumn{2}{c} {\large $f_{r}(r)=\int_{r-R_{2}}^{R_{1}} g_{2}(\rho,r)f_{\rho}(\rho) \,d\rho$} \\
    \hline \hline
\end{NiceTabular}
}
\end{table*}

The conditional PDF $f_{r|\rho}(r|\rho)$ provides direct insight into boundary effects: the transition at $r = R_2 - \rho$ separates the regime where the ring/shell around node~1 is fully contained in the outer region from the regime where it is truncated; as $\rho$ increases toward the boundary, the fully contained interval shrinks and truncation becomes dominant, reducing the likelihood of small $r$ and shifting probability mass toward larger distances. 

To compute $f_{\mathbf{r}/\bm{\rho}}(r/\rho)$, it is assumed that node 2 is located on a circular ring or spherical shell of radius $r$ centered at node 1. Only the portion of this ring or shell that is enclosed in the disk or sphere of radius $R_2$ contributes to the conditional probability. Depending on the relative values of $r$, $\rho$, $R_1$, and $R_2$, this region may be fully contained within, partially overlap with, or lie outside the corresponding disk or sphere. 

Table~\ref{Table1} specifies the three intervals of $r$ over which the conditional probability takes distinct forms, along with the corresponding integration limits in \eqref{CondProb}. The three intervals are defined as follows.

\begin{itemize}
    \item \textbf{Interval I} ($0 < r < R_2-R_1$): The circular ring or spherical shell (dashed line in Table~\ref{Table1}) is fully contained within the disk/sphere of radius $R_2$ for all $\rho$. Therefore, $g_1(\rho, r)$ is used\footnote{\label{note}As defined in Table~\ref{Table2} for 2D and Table~\ref{Table3} for 3D.}.
       
    \item \textbf{Interval II} ($R_2-R_1 < r < R_{2}$): The ring or shell is fully contained within the disk/sphere for $0 < \rho < R_{2} - r$ and partially intersects it for $R_{2} - r < \rho < R_{1}$. Both $g_1(\rho, r)$ and $g_2(\rho, r)$ are used\footref{note}. 
        
    \item \textbf{Interval III} ($R_{2} < r < R_1+R_2$): The ring or shell lies outside the disk/sphere for $0 < \rho < r - R_{2}$ (probability = 0) and partially intersects it for $r - R_{2} < \rho < R_{1}$. The function $g_2(\rho, r)$ is used\footref{note}.
\end{itemize}

The auxiliary functions $g_1(\rho,r)$ and $g_2(\rho,r)$ have a direct geometric interpretation: they represent the conditional contribution of the circular ring (2D) or spherical shell (3D) of radius $r$ centered at node~1 when it is fully contained inside the outer region ($r < R_2-\rho$) or when it is partially truncated by the boundary ($r \ge R_2-\rho$), respectively.  The explicit expressions of these functions, along with $f_{\rho}(\rho)$ are given in Table \ref{Table2} for the 2D case and in Table \ref{Table3} for the 3D case.

\begin{table}[h]
\centering
\caption{Definition of $g_{1}(\rho, r), g_{2}(\rho, r)$ and $f_{\rho}(\rho)$ for the 2D case.}
\label{Table2} 
\resizebox{\linewidth}{!}{
\renewcommand{\arraystretch}{1.8}
\begin{tabular}{c|c|c} 
 \hline \hline
 \textbf{Function} & \textbf{s1, s4} & \textbf{s2, s3} \\ [0.5ex] 
 \hline
$g_{1}(\rho, r)$ & $\dfrac{2r}{R_{2}^{2}}$ & $-\dfrac{4rh_{2}(\rho, r)}{R_{2}^{2}}$\\ [1.5ex] 
\hline
$g_{2}(\rho, r)$ & $\dfrac{2r\arccos\left(\frac{h_{2}(\rho, r)}{2\rho r}\right)}{\pi R_{2}^{2}}$ & $\dfrac{4r(h_{1}(\rho, r)-h_{2}(\rho, r)\arccos(\frac{h_{2}(\rho, r)}{2\rho r}))}{\pi R_{2}^{4}}$ \\ [1.5ex] 
\hline
$h_{1}(\rho, r)$ & - & $\sqrt{(r^{2}-(R_{2}-\rho)^{2})((R_{2}+\rho)^{2}-r^{2})}$ \\ [1.5ex] 
\hline
$h_{2}(\rho, r)$ & \multicolumn{2}{|c} {$\rho^{2}+r^{2}-R_{2}^{2}$} \\ [1.5ex] 
\hline\hline
 \textbf{Function} & \textbf{s1, s3} & \textbf{s2, s4} \\ [0.5ex] 
 \hline
 $f_{\rho}(\rho)$ & $\dfrac{4\rho}{R_{1}^{2}}\left(1-\dfrac{\rho^{2}}{R_{1}^{2}} \right)$ & $\dfrac{2\rho}{R_{1}^{2}}$ \\ [1.5ex] 
\hline \hline
\end{tabular}
}
\end{table}

\section{Internodal Distance Distributions}
\label{sec:distribution}
\subsection{2D Networks}

\begin{theorem}
\label{pdf2Da} The PDF expression for the internodal distance distribution, $r$, in a 2D network with $R_{1}<R_{2}$, is given by
\begin{eqnarray}
\hspace{-8pt}f_{\mathbf{r}}(r)\hspace{-3pt} =\hspace{-3pt}
\begin{cases}
\frac{2rq_{1}(r)}{R_{2}^{2}}, & 0\leq r <R_2-R_1, \\
\frac{r}{2\pi R_{1}^{4} R_{2}^{2}} 
\Bigg( q_{2}(r) \arccos \left(\frac{k_{1}(r)}{2rR_{1}}\right) 
+ \\ 
q_{3}(r) \arccos \left(\frac{k_{2}(r)}{2rR_{2}}\right) + q_{4}(r) \hspace{-4pt}\Bigg), 
& \hspace{-16pt}R_2\hspace{-3pt}-\hspace{-3pt}R_1 \leq r \leq \hspace{-3pt}R_1\hspace{-3pt}+\hspace{-2pt}R_2,
\end{cases}
\hspace{-30pt}\label{Gen1}
\end{eqnarray}
where $k_{1}(r)=r^{2}+R_{1}^{2}-R_{2}^{2}$ and $k_{2}(r)=r^{2}-R_{1}^{2}+R_{2}^{2}$ are system dependent parameters and the polynomials $q_{1}(r)-q_{4}(r)$ are specified for each Scenario and tabulated in Table \ref{qTable}.
\end{theorem}

\begin{table}[h]
\centering
\caption{Definition of $g_{1}(\rho, r), g_{2}(\rho, r)$ and $f_{\rho}(\rho)$ for the 3D Case.}
\label{Table3} 
\resizebox{\linewidth}{!}{\renewcommand{\arraystretch}{1.5}
\begin{tabular}{c|c|c} 
 \hline \hline
 \textbf{Function} & \textbf{s1, s4} & \textbf{s2, s3} \\ [0.5ex] 
 \hline
$g_{1}(\rho, r)$ & $\dfrac{3r^2}{R_{2}^{3}}$ & $\dfrac{35r^2\left(104\rho^2r^2+6h_{3}(\rho, r)h_{4}(\rho, r) \right)}{432R_{2}^{7}}$ \\ [1.5ex] 
\hline
$g_{2}(\rho, r)$ & $\dfrac{3r^2}{2R_{2}^3}\left(1-\dfrac{h_{3}(\rho, r)}{2\rho r }\right)$ & $\dfrac{35r(25R_{2}^2-13(r-\rho)^2)h_{5}(\rho, r)}{864R_{2}^7\rho}$ \\ [1.5ex] 
\hline
$h_{3}(\rho, r)$ & \multicolumn{2}{|c} {$\rho^{2}+r^{2}-R_{2}^{2}$} \\ [1.5ex] 
\hline
$h_{4}(\rho, r)$ & - & $13r^2-21R_{2}^2+13\rho^2$ \\ [1.5ex] 
\hline
$h_{5}(\rho, r)$ & - & $(r-\rho+R_{2})^2(\rho-r+R_{2})^2$ \\ [1.5ex] 
\hline\hline
 \textbf{Function} & \textbf{s1, s3} & \textbf{s2, s4} \\ [0.5ex] 
 \hline
 $f_{\rho}(\rho)$ & $\dfrac{35}{72}\left(\dfrac{21 \rho^2}{R_{1}^3}-\dfrac{34 \rho^4}{R_{1}^5}+\dfrac{13 \rho^6}{R_{1}^7} \right)$ & $\dfrac{3\rho^2}{R_{1}^{3}}$ \\ [1.5ex] 
\hline \hline
\end{tabular}
}
\end{table}

\begin{table}[h]
\centering
\caption{Definition of $q_{1}(r) - q_{4}(r)$ for the 2D case.}
\label{qTable} 
\resizebox{\linewidth}{!}{
\renewcommand{\arraystretch}{1.7}
\begin{tabular}{c|c|c}
 \hline \hline
 & \textbf{s1} & \textbf{s3} \\ [0.5ex] 
 \hline
$q_{1}(r)$ & $1$ & $-\dfrac{2(3r^{2}+R_{1}^{2}-3R_{2}^{2})}{3R_{2}^{2}}$  \\  [1.5ex] 
\hline
$q_{2}(r)$ & $4R_{1}^{4}$ & $-\dfrac{24R_{1}^{4}(3r^{2}+R_{1}^{2}-3R_{2}^{2})}{9R_{2}^{2}}$ \\
\hline
$q_{3}(r)$ & $4R_{2}^{2}(2R_{1}^{2}-2r^{2}-R_{2}^{2})$ & $-\dfrac{24R_{2}^{2}(3r^{2}+R_{2}^{2}-3R_{1}^{2})}{9}$ \\
  [0.5ex] 
\hline
$q_{4}(r)$ & $(r^{2}\!-\!3R_{1}^{2}\!+\!5R_{2}^{2})h_{1}(R_{1},r)$ & 
$\dfrac{2h_{1}(R_{1},r)}{9R_{2}^{2}}\times$\\[-4pt]& &$(-r^{4}\!+\!8r^{2}\left(R_{1}^{2}\!+\!R_{2}^{2})\!+\!17(R_{1}^{4}\!+\!R_{2}^{4})\!-\!22R_{1}^{2}R_{2}^{2}\right)$ \\[1.5ex]
\hline
\hline
 &\textbf{s2} & \textbf{s4}\\  [0.5ex] 
\hline
$q_{1}(r)$ & $-\dfrac{2r^{2}+R_{1}^{2}-2R_{2}^{2}}{R_{2}^{2}}$ & $1$\\  [0.5ex] 
\hline
$q_{2}(r)$ & $-\dfrac{4R_{1}^{4}(2r^{2}+R_{1}^{2}-2R_{2}^{2})}{R_{2}^{2}}$ & $4R_{1}^{4}$ \\  [1.5ex] 
\hline
$q_{3}(r)$ & $4R_{1}^{2}R_{2}^{2}$ & $4R_{1}^{2}R_{2}^{2}$\\  [0.5ex] 
\hline
$q_{4}(r)$ & $\dfrac{R_{1}^{2}(r^{2}\!+\!5R_{1}^{2}\!-\!3R_{2}^{2})h_{1}(R_{1},r)}{R_{2}^{2}}$ & $-4R_{1}^{2}\left(\dfrac{R_{1}k_{1}(r)}{2r}\sqrt{1-\left(\dfrac{k_{1}(r)}{2rR_{1}}\right)^{2}}\right.$ \\ 
 & & $\left. + \dfrac{R_{2}k_{2}(r)}{2r}\sqrt{1-\left(\dfrac{k_{2}(r)}{2rR_{2}}\right)^{2}}\right)$\\  [1.5ex] 
\hline \hline
\end{tabular}
}
\end{table}

\begin{table}[h]
\centering
\caption{Definition of $s_{1}(r)$ and $s_{2}(r)$ for the 2D case.}
\label{sTable} 
\resizebox{\linewidth}{!}{
\renewcommand{\arraystretch}{1}
\scriptsize
\scalebox{0.85}{
\begin{tabular}{c|c|c|c} 
 \hline \hline
  & \textbf{s1, s2} & \textbf{s3} & \textbf{s4} \\ [0.5ex] 
 \hline
$s_{1}(r)$ & $\frac{2(-r^{2}+R^{2})}{\pi}$ &  $\frac{4(-3r^{2}+2R^{2})}{3\pi}$ & $\frac{2R^{2}}{\pi}$\\ [1.2ex] 
\hline
$s_{2}(r)$ & $\frac{r(r^{2}+2R^{2})}{2\pi R}$  & $\frac{-r^{5}+16r^{3}R^{2}+12rR^{4}}{9\pi R^{3}}$ & $-\frac{rR}{\pi}$\\ [1.2ex] 
\hline \hline
\end{tabular}}
}
\end{table}

\begin{IEEEproof}  The proof is provided in Appendix A. 
\end{IEEEproof}

\begin{corollary}
\label{pdf2Db} The PDF of $r$ for $R_{1}=R_{2}=R$ is given by
\begin{equation}
f_{\mathbf{r}}(r)\hspace{-2pt}=\hspace{-2pt}\dfrac{2r\hspace{-2pt}\left(\hspace{-2pt}s_{1}(r)\arccos\hspace{-2pt}\left(\frac{r}{2R}\right)\hspace{-2pt}+\hspace{-2pt}s_{2}(r)\sqrt{1-{(\frac{r}{2R}})^{2}}\right)}{R^{4}}, 0\leq r<2R ,
\label{pdfr}
\end{equation}
where $s_{1}(r)$ and $s_{2}(r)$ are listed in Table \ref{sTable} for each scenario.
\end{corollary}
\begin{IEEEproof}
The result follows by setting $R_1 = R_2 = R$ in \eqref{Gen1}.
\end{IEEEproof}

\subsection{3D Networks}
\begin{theorem}
\label{pdf3Da} 
The PDF expression for the internodal distance distribution, $r$, in a 3D network with $R_{1}<R_{2}$, is given by 
\begin{equation}
f_{\mathbf{r}}(r)=\left\{
\begin{array}{ll}
\displaystyle\sum_{n=1}^{3} a_{2n}r^{2n}, & 0\leq r<
R_2-R_1, \\
& \\
\displaystyle\sum_{n=1}^{13} b_{n}r^{n} , & R_2-R_1
\leq r\leq R_1+R_2, \\
\end{array} 
\right. \label{Gen4}
\end{equation}
where $a_{2n}$ and $b_{n}$ are listed in Tables \ref{aTable}, \ref{b_main} and \ref{b_s3} for each scenario, whereas $b_{8}=b_{10}=b_{12}=0$.
\end{theorem}

\begin{table}[h]
\centering
\caption{Definition of $a_{2n}$ for the 3D case.}
\label{aTable} 
\resizebox{\linewidth}{!}{\renewcommand{\arraystretch}{1.5}
\begin{tabular}{c|c|c|c} 
 \hline \hline
 \cline{2-4}
  & \textbf{s1, s4} & \textbf{s2} & \textbf{s3} \\ [0.5ex] 
 \hline
$a_{2}$ & $\dfrac{3}{R_{2}^{3}}$ & $\dfrac{65R_{1}^4-238R_{1}^2R_{2}^2+245R_{2}^4}{24R_{2}^7}$ & $\dfrac{35(2275R_{1}^4-11594R_{1}^2R_{2}^2+18711R_{2}^4)}{64152R_{2}^7}$ \\  [1.5ex] 
\hline
$a_{4}$ & $0$ & $\dfrac{35(13R_{1}^2-17R_{2}^2)}{36R_{2}^7}$ & $\dfrac{35(2015R_{1}^2-4131R_{2}^2)}{8748R_{2}^7}$ \\  [1.5ex] 
\hline
$a_{6}$ & $0$ & \multicolumn{2}{|c}{$\dfrac{455}{72R_{2}^7}$} \\  [1.5ex] 
\hline \hline
\end{tabular}
}
\end{table}

\begin{IEEEproof}  The proof is provided in Appendix B. 
\end{IEEEproof}

\begin{corollary}
\label{pdf3Db} The PDF of $r$ for $R_1 = R_2 = R$ is given by
\begin{equation}
f_{\mathbf{r}}(r)=\displaystyle\sum_{n=2}^{13} c_{n}r^{n}, 0\leq r<2R ,
\label{Gen2}
\end{equation}
where $c_{n}$ are listed in Table \ref{cTable} for each scenario, whereas $c_{8}=c_{10}=c_{12}=0$.
\end{corollary}

\begin{IEEEproof}
The result follows by setting $R_1\hspace{-2pt}=R_2=R$ in \eqref{Gen4}.
\end{IEEEproof}

 The coefficients appearing in Theorems~1 and~2 result from integrating these geometric terms with the radial spatial density $f_\rho(\rho)$ and thus capture the combined effect of geometry and mobility. The remaining auxiliary functions are employed to make the final expressions more compact.

\subsection{Approximation with Beta Distributions}
\label{sec:beta}
While the exact PDFs are given in closed form, tractable adoption in higher-layer analyses can be achieved by approximating them using a beta distribution, with PDF \cite{J:Ermolova}

\begin{equation}
{f_{\bf{x}}}(x) = \frac{{{x^{\alpha  - 1}}{{(1 - x)}^{\beta  - 1}}}}{{B(\alpha ,\beta )}},\begin{array}{*{20}{c}}
{}&{}&{x \in \left[ {0,1} \right]},
\end{array}
\label{betaDist}
\end{equation} 
where $\alpha$ and $\beta$ are the distribution parameters, and $B(\alpha,\beta)$ is the beta function. The parameters used for the approximation are obtained by matching the first and second moments of the target distribution with those of the beta distribution \cite{J:Ermolova}. Solving the resulting system yields the corresponding optimal parameter values.

\begin{figure}[h]
\centering
\begin{minipage}{0.23\textwidth}
\centering
\includegraphics[width=\linewidth]{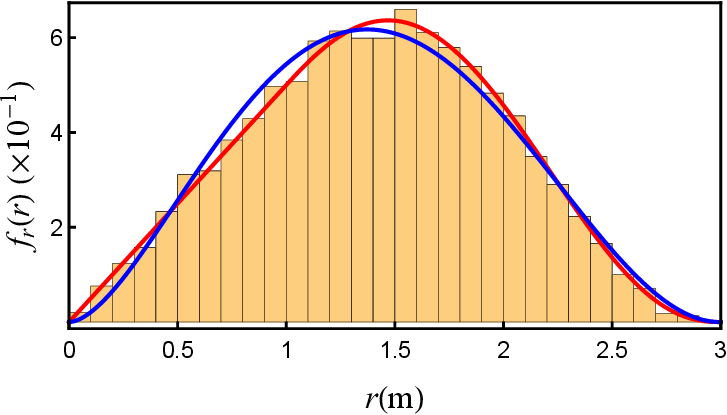}\\
{\scriptsize (a) s1, $\{\alpha,\beta\}=\{2.753,3.080\}$}
\end{minipage}\hfill
\begin{minipage}{0.23\textwidth}
\centering
\includegraphics[width=\linewidth]{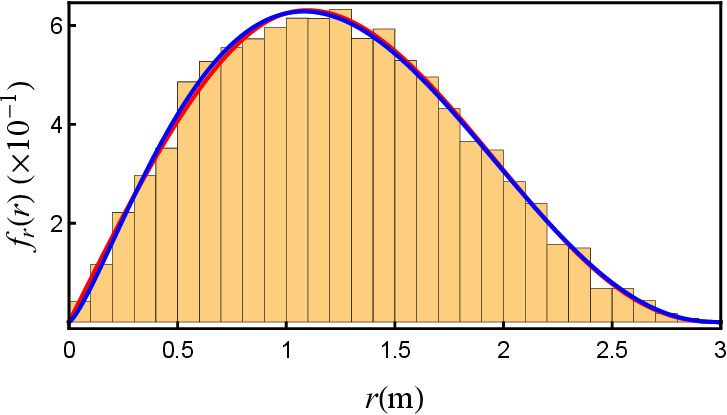}\\ 
{\scriptsize (b) s2, $\{\alpha,\beta\}=\{2.333,3.366\}$}
\end{minipage}

\vspace{0.5em}

\begin{minipage}{0.23\textwidth}
\centering
\includegraphics[width=\linewidth]{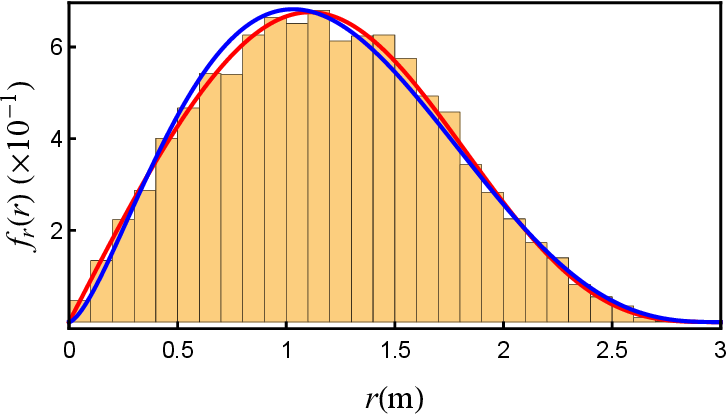}\\
{\scriptsize (c) s3, $\{\alpha,\beta\}=\{2.550,3.960\}$}
\end{minipage}\hfill
\begin{minipage}{0.23\textwidth}
\centering
\includegraphics[width=\linewidth]{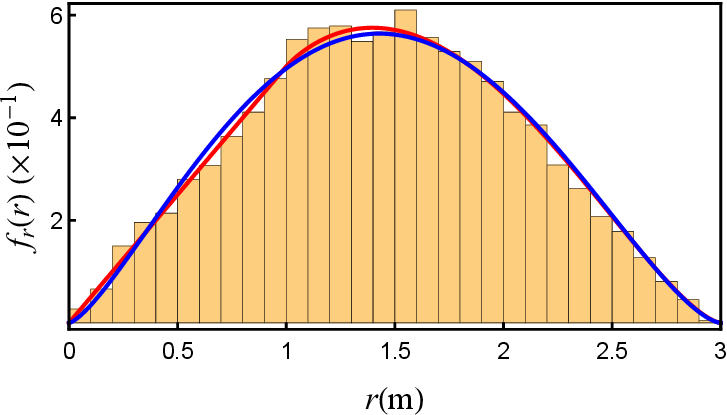}\\
{\scriptsize (d) s4, $\{\alpha,\beta\}=\{2.410,2.552\}$}
\end{minipage}
\caption{PDF for the 2D case with $\{R_1,R_2\}=\{1,2\}\text{m}$: analytical solutions (red), beta approximations (blue), and Monte Carlo results (histogram).}
\label{Figure2}
\end{figure}

\begin{figure}[h]
\centering
\begin{minipage}{0.23\textwidth}
\centering
\includegraphics[width=\linewidth]{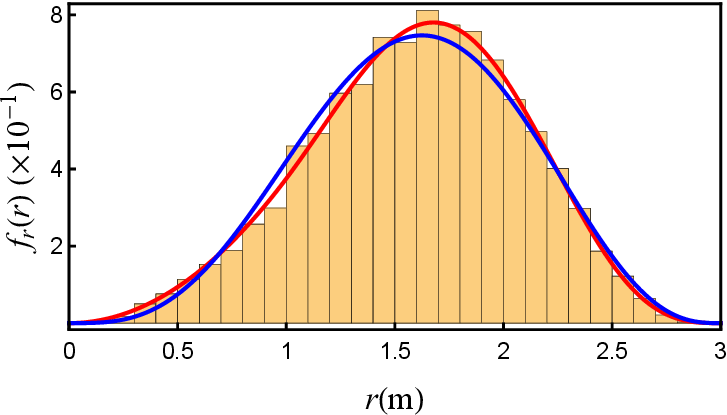}\\
{\scriptsize (a) s1, $\{\alpha,\beta\}=\{4.422,3.898\}$}
\end{minipage}\hfill
\begin{minipage}{0.23\textwidth}
\centering
\includegraphics[width=\linewidth]{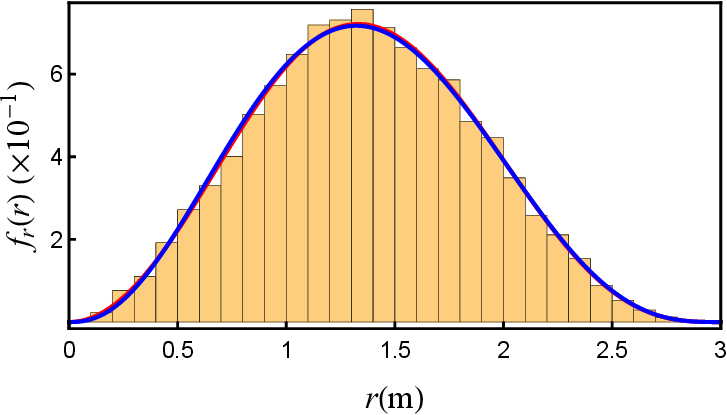}\\
{\scriptsize (b) s2, $\{\alpha,\beta\}=\{3.495,4.166\}$}
\end{minipage}

\vspace{0.5em}

\begin{minipage}{0.23\textwidth}
\centering
\includegraphics[width=\linewidth]{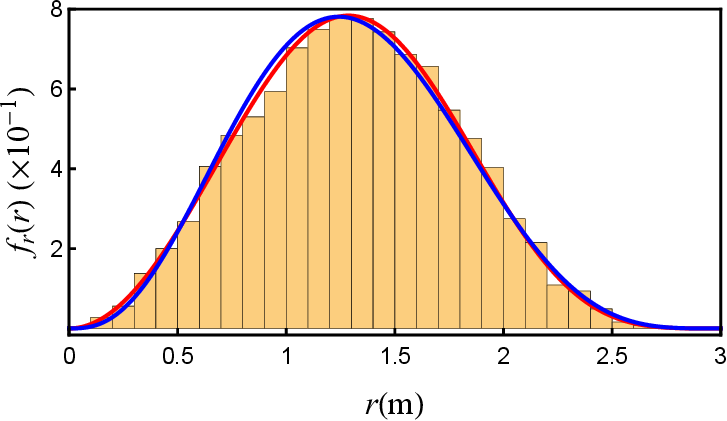}\\
{\scriptsize (c) s3, $\{\alpha,\beta\}=\{3.846,5.030\}$}
\end{minipage}\hfill
\begin{minipage}{0.23\textwidth}
\centering
\includegraphics[width=\linewidth]{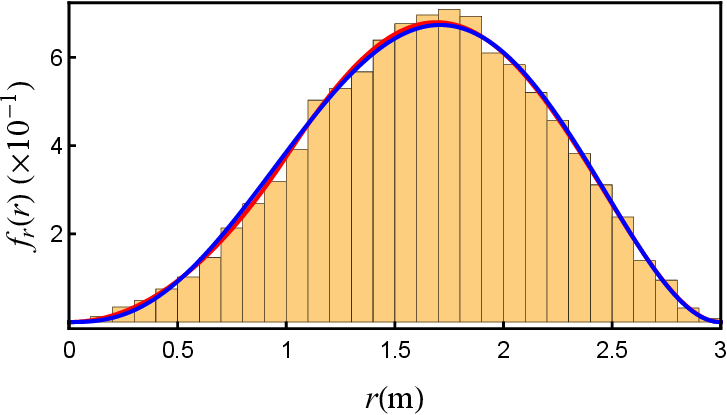}\\
{\scriptsize (d) s4, $\{\alpha,\beta\}=\{3.724,3.058\}$}
\end{minipage}
\caption{PDF plots for the 3D case with $\{R_1,R_2\}=\{1,2\}\text{m}$: analytical solutions (red), beta approximations (blue), and Monte Carlo results (histogram).}
\label{Figure3}
\end{figure}

 Figures \ref{Figure2} and \ref{Figure3} illustrate representative PDF plots for the 2D and 3D cases with $R_{2}/R_{1}=2$, alongside Monte Carlo results of $100.000$ independent realizations. The approximation accuracy is only negligibly affected by small perturbations in $\alpha$ and $\beta$, which arises solely from truncation or rounding errors.

\begin{table*}[t]
\centering
\caption{Definition of coefficients $b_n$ for s1, s2, and s4 (3D case).}
\label{b_main}
\scriptsize
\renewcommand{\arraystretch}{1.5}
\scalebox{0.9}{
\begin{tabular}{c|c|c|c}
\hline \hline
 & \textbf{s1} & \textbf{s2} & \textbf{s4} \\ 
\hline
$b_{1}$ & $\dfrac{35(29R_{1}^2-13R_{2}^2)(R_{2}^2-R_{1}^2)^3}{2304R_{1}^7R_{2}^3}$ & $\dfrac{35(13R_{1}^2-29R_{2}^2)(R_{2}^2-R_{1}^2)^3}{2304R_{2}^7R_{1}^3}$ & $-\dfrac{9(R_{2}^2-R_{1}^2)^2}{16R_{1}^3R_{2}^3}$ \\ 
\hline
$b_{2}$ & $\dfrac{72R_{1}^7+245R_{1}^4R_{2}^3-238R_{1}^2R_{2}^5+65R_{2}^7}{48R_{1}^7R_{2}^3}$ & $\dfrac{72R_{2}^7+245R_{2}^4R_{1}^3-238R_{2}^2R_{1}^5+65R_{1}^7}{48R_{2}^7R_{1}^3}$ & $\dfrac{3(R_{1}^3+R_{2}^3)}{2R_{1}^3R_{2}^3}$ \\
\hline
$b_{3}$ & $\dfrac{35(R_{2}^2-R_{1}^2)(25R_{1}^4+88R_{1}^2R_{2}^2-65R_{2}^4)}{576R_{1}^7R_{2}^3}$ & $\dfrac{35(R_{2}^2-R_{1}^2)(65R_{1}^4-25R_{2}^4-88R_{2}^2R_{1}^2)}{576R_{2}^7R_{1}^3}$ & $-\dfrac{9(R_{1}^2+R_{2}^2)}{8R_{1}^3R_{2}^3}$ \\
\hline
$b_{4}$ & $\dfrac{35(13R_{2}^2-17R_{1}^2)}{72R_{1}^7}$ & $\dfrac{35(13R_{1}^2-17R_{2}^2)}{72R_{2}^7}$ & $0$ \\
\hline
$b_{5}$ & $\dfrac{35(7R_{1}^4+34R_{1}^2R_{2}^2-65R_{2}^4)}{384R_{1}^7R_{2}^3}$ & $\dfrac{35(7R_{2}^4+34R_{2}^2R_{1}^2-65R_{1}^4)}{384R_{2}^7R_{1}^3}$ & $\dfrac{3}{16R_{1}^3R_{2}^3}$ \\
\hline
$b_{6}$ & $\dfrac{455}{144R_{1}^7}$ & $\dfrac{455}{144R_{2}^7}$ & $0$ \\
\hline
$b_{7}$ & $-\dfrac{7(17R_{1}^2+65R_{2}^2)}{576R_{1}^7R_{2}^3}$ & $-\dfrac{7(17R_{2}^2+65R_{1}^2)}{576R_{2}^7R_{1}^3}$ & $0$ \\
\hline
$b_{9}$ & $\dfrac{65}{2304R_{1}^7R_{2}^3}$ & $\dfrac{65}{2304R_{2}^7R_{1}^3}$ & $0$ \\
\hline
$b_{11},b_{13}$ & $0$ & $0$ & $0$ \\
\hline \hline
\end{tabular}}
\end{table*}

\begin{table}[h]
\centering
\caption{Definition of coefficients $b_n$ for s3 (3D case).}
\label{b_s3}
\scriptsize
\renewcommand{\arraystretch}{2}
\scalebox{0.9}{
\begin{tabular}{c|c}
\hline \hline
& \textbf{s3} \\ 
\hline
$b_{1}$ & $\dfrac{245(R_{2}^2-R_{1}^2)^4(1442R_{1}^2R_{2}^2-481(R_{1}^4+R_{2}^4))}{1492992R_{1}^7R_{2}^7}$ \\
\hline
$b_{2}$ & $\frac{35(2275(R_{1}^{11}+R_{2}^{11})-11594(R_{1}^9R_{2}^2+R_{1}^2R_{2}^9)+18711(R_{1}^7R_{2}^4+R_{1}^4R_{2}^7))}{128304R_{1}^7R_{2}^7}$ \\
\hline
$b_{3}$ & $\dfrac{1225(R_{2}^2-R_{1}^2)^2(R_{1}^2+R_{2}^2)(350R_{1}^2R_{2}^2-143(R_{1}^4+R_{2}^4))}{82944R_{1}^7R_{2}^7}$ \\
\hline
$b_{4}$ & $\dfrac{35(2015(R_{1}^9+R_{2}^9)-4131(R_{1}^7R_{2}^2+R_{1}^2R_{2}^7))}{17496R_{1}^7R_{2}^7}$ \\
\hline
$b_{5}$ & $\dfrac{1225(1700(R_{1}^6R_{2}^2+R_{1}^2R_{2}^6)+882R_{1}^4R_{2}^4-1885(R_{1}^8+R_{2}^8))}{497664R_{1}^7R_{2}^7}$ \\
\hline
$b_{6}$ & $\dfrac{455(R_{1}^7+R_{2}^7)}{144R_{1}^7R_{2}^7}$ \\
\hline
$b_{7}$ & $\dfrac{245(R_{1}^2+R_{2}^2)(554R_{1}^2R_{2}^2-1625(R_{1}^4+R_{2}^4))}{373248R_{1}^7R_{2}^7}$ \\
\hline
$b_{9}$ & $\dfrac{35(455(R_{1}^4+R_{2}^4)+578R_{1}^2R_{2}^2)}{165888R_{1}^7R_{2}^7}$ \\
\hline
$b_{11}$ & $-\dfrac{7735(R_{1}^2+R_{2}^2)}{746496R_{1}^7R_{2}^7}$ \\
\hline
$b_{13}$ & $\dfrac{29575}{49268736R_{1}^7R_{2}^7}$ \\
\hline \hline
\end{tabular}}
\end{table}

\begin{table}[h]
\begin{center}
\caption{Definition of $c_{n}$ for the 3D case.}
\label{cTable} 
\small \renewcommand{\arraystretch}{1.2}
\scalebox{0.9}{
\begin{tabular}{c|c|c|c|c} 
 \hline \hline
 & \multicolumn{2}{|c|}{\textbf{s1, s2}} & \textbf{s3} & \textbf{s4} \\ [0.5ex] 
 \hline
$c_{2}$ & \multicolumn{2}{|c|}{$\frac{3}{R^{3}}$}  & $\frac{41090}{8019R^{3}}$ & $\frac{3}{R^{3}}$\\ [1ex] 
\hline
$c_{3}$ & \multicolumn{3}{|c|}{$0$} & $-\frac{9}{4R^{4}}$\\ [1ex] 
\hline
$c_{4}$ & \multicolumn{2}{|c|}{$-\frac{35}{18R^{5}}$}  & $-\frac{18515}{2187R^{5}}$ & $0$\\ [1ex] 
\hline
$c_{5}$ & \multicolumn{2}{|c|}{$-\frac{35}{16R^{6}}$}  & $\frac{1225}{972R^{6}}$ & $\frac{3}{16R^{6}}$\\ [1ex] 
\hline
$c_{6}$ & \multicolumn{2}{|c|}{$\frac{455}{144R^{7}}$}  & $\frac{455}{72R^{7}}$ & $0$\\ [1ex] 
\hline
$c_{7}$ & \multicolumn{2}{|c|}{$-\frac{287}{288R^{8}}$}  & $-\frac{82565}{23328R^{8}}$ & $0$\\ [1ex] 
\hline
$c_{9}$ & \multicolumn{2}{|c|}{$\frac{65}{2304R^{10}}$}  & $\frac{1085}{3456R^{10}}$ & $0$\\ [1ex] 
\hline
$c_{11}$ & \multicolumn{2}{|c|}{$0$} & $-\frac{7735}{373248R^{12}}$ & $0$\\ [1ex]
\hline
$c_{13}$ & \multicolumn{2}{|c|}{$0$} & $\frac{29575}{49268736R^{14}}$ & $0$\\ [1ex]  
\hline \hline
\end{tabular}}
\end{center}
\end{table}

\section{Concluding remarks}
This letter presents a unified framework for internodal distance PDFs in 2D/3D networks with nodes in concentric circular or spherical regions, covering all node-pair types under uniform and RWP models, including equal-radius cases. In the 2D case, the novelty lies in the unified treatment of all static/mobile node-pair combinations under both homogeneous and heterogeneous region sizes, as well as mobile/mobile node pairs under heterogeneous region sizes. While this work focuses on the analytical characterization of internodal distance distributions, these results are primarily intended as building blocks for the evaluation of communication-oriented performance metrics such as SNR/SINR, coverage probability, outage probability, interference, and connectivity. 

 Future work may extend the proposed framework to alternative mobility models and non-concentric or irregular deployment regions. Additionally, explicitly accounting for temporal correlation and distance evolution is of interest.
 
 \appendices

\section{Proof of Theorem 1}
\setcounter{equation}{0} \renewcommand\theequation{A.\arabic{equation}}
\indent$\bullet~$\textbf{s2 and s3}: The joint PDF of node 2's position under the RWP model, within a circular disk of radius $R_2$, centered at $(\rho,0)$, with respect to a reference point at $(0,0)$, is expressed in polar coordinates as
\begin{equation}
f_{\mathbf{r},\bm{\vartheta}}(r,\vartheta)=\frac{2r}{\pi R_{2}^2}\left(1-\frac{r^2+{\rho}^2 -2r{\rho}{\cos}\vartheta}{R_{2}^2}\right).
\label{Gen2D2}
\end{equation}
Considering the general equation of a circle in polar coordinates centered on $(\rho,0)$, the angle $\vartheta$ can be obtained as
\begin{equation}
\vartheta(r)=\arccos\left(\frac{r^{2}-R_{2}^{2}+{\rho}^2}{2r{\rho}} \right).
\label{Gen2D3}
\end{equation}
Consequently, the conditional PDF of the internodal distance, given that node 1 is located at $(0,0)$, is obtained by integrating \eqref{Gen2D2}  over $\vartheta(r)$ in two distinct branches. In the first branch, integration is performed over the interval $[0, 2\pi]$, whereas in the second branch, it is carried out over $[- \vartheta(r), \vartheta(r)]$. The final expression can be obtained as
\begin{equation}
f_{\mathbf{r}/\bm{\rho}}(r/\rho)=\left\{
\begin{array}{ll}
g_{1}(\rho,r), & 0\leq r<
R_{2}-{\rho} \\
& \\
 g_{2}(\rho,r), & R_{2}-{\rho}
\leq r\leq R_{2}+{\rho}, \\
\end{array} 
\right.
\label{Gen2D4}
\end{equation}
where $g_{1}(\rho,r)$, $g_{2}(\rho,r)$ are presented in Table \ref{Table2}. \\ 
\indent $\bullet~$\textbf{s1 and s4}: The joint PDF remains invariant to the displacement of the center when node~2 is uniformly distributed within a circular region of radius $R_{2}$. That is\footnote{\label{note1}Expressed as a joint PDF to include the Jacobian factor; independence from angle follows from symmetry.},
\begin{equation}
f_{\mathbf{r},\bm{\vartheta}}(r,\vartheta)=\dfrac{r}{\pi R_{2}^2}.
\label{Gen2D5}
\end{equation}
The conditional PDF of $r$ is derived by integrating \eqref{Gen2D5}  over $\vartheta(r)$ in two branches: $[0, 2\pi]$ for the first, and $[- \vartheta(r), \vartheta(r)]$ for the second.
Finally, an equivalent expression to \eqref{Gen2D4}  is derived, where $g_1(\rho,r)$ and $g_2(\rho,r)$ take distinct values, as shown in Table \ref{Table2}.
 Note that the last two branches (Interval II, III) yield the same result.

\section{Proof of Theorem 2}
\setcounter{equation}{0} \renewcommand\theequation{B.\arabic{equation}}
\indent$\bullet~$\textbf{s2 and s3}: The joint PDF for node 2's position under the RWP model, within a sphere of radius $R_2$ centered at $(0,0,\rho)$, with respect to a reference point at $(0,0,0)$, is expressed in spherical coordinates as\footref{note1}
\begin{align}
    f_{\mathbf{r},\bm{\vartheta},\bm{\varphi}}(r,\vartheta,\varphi) &= 
    \frac{35r^2\sin\vartheta}{288\pi} \Bigg( \frac{21}{R_{2}^3} 
    - \frac{34\left(r^2 + {\rho}^2 - 2r{\rho} \cos\vartheta\right)}{R_{2}^5} \notag \\
    &\quad+ \frac{13\left(r^2 + {\rho}^2 - 2r{\rho} \cos\vartheta\right)^2}{R_{2}^7} \Bigg).
    \label{Gen3D2}
\end{align}
Considering the general equation of a sphere centered at $(0,0,\rho)$,  the angle $\vartheta(r)$ can be obtained as in \eqref{Gen2D3} . The conditional PDF of $r$, given that node 1 is located at $(0,0,0)$ is obtained by integrating $f_{r,\vartheta,\varphi}(r,\vartheta,\varphi)$ over $\varphi \in [0, 2\pi]$, and $\vartheta$ in two branches: $[0, \pi]$ for the first, and $[0, \vartheta(r)]$ for the second.
The resulting expression retains the same form as \eqref{Gen2D4} ; however, in this case, the functions $g_{1}(\rho, r)$ and $g_{2}(\rho, r)$ are provided in Table~\ref{Table3}.\\ 
\indent$\bullet~$\textbf{s1 and s4}: The joint PDF remains invariant to the displacement of the center when node~2 is uniformly distributed within a sphere with radius $R_{2}$. That is\footref{note1},

\begin{equation}
f_{\mathbf{r},\bm{\vartheta},\bm{\varphi}}(r,\vartheta,\varphi)=\frac{3r^2\sin\vartheta}{4\pi R_{2}^3}.
\label{Gen5}
\end{equation}
The conditional PDF of the internodal distance is obtained by integrating \eqref{Gen5}  over $\vartheta(r)$ in two branches: $[0, \pi]$ for the first, and $[0,\vartheta(r)]$ for the second. Finally, an equivalent mathematical expression to \eqref{Gen2D4} is derived, where $g_{1}(\rho,r)$ and $g_{2}(\rho,r)$ take distinct values, as shown in Table \ref{Table3}. Note that the last two branches (Interval II, III) yield the same result.

\bibliographystyle{IEEEtran}
\bibliography{IEEEabrv,References}

\begin{thebibliography}{10}
\providecommand{\url}[1]{#1}
\csname url@samestyle\endcsname
\providecommand{\newblock}{\relax}
\providecommand{\bibinfo}[2]{#2}
\providecommand{\BIBentrySTDinterwordspacing}{\spaceskip=0pt\relax}
\providecommand{\BIBentryALTinterwordstretchfactor}{4}
\providecommand{\BIBentryALTinterwordspacing}{\spaceskip=\fontdimen2\font plus
\BIBentryALTinterwordstretchfactor\fontdimen3\font minus \fontdimen4\font\relax}
\providecommand{\BIBforeignlanguage}[2]{{%
\expandafter\ifx\csname l@#1\endcsname\relax
\typeout{** WARNING: IEEEtran.bst: No hyphenation pattern has been}%
\typeout{** loaded for the language `#1'. Using the pattern for}%
\typeout{** the default language instead.}%
\else
\language=\csname l@#1\endcsname
\fi
#2}}
\providecommand{\BIBdecl}{\relax}
\BIBdecl

\bibitem{J:Moltchanov}
D.~Moltchanov, ``Distance distributions in random networks,'' \emph{Ad Hoc Netw.}, vol.~10, no.~6, pp. 1146--1166, 2012.

\bibitem{B:Haenggi}
M.~Haenggi, \emph{Stochastic Geometry for Wireless Networks}.\hskip 1em plus 0.5em minus 0.4em\relax New York, USA: Cambridge Univ. Press, 2012.

\bibitem{J:Badarneh}
O.~S. Badarneh, M.~K. Awad, S.~Muhaidat, and F.~S. Almehmadi, ``Performance analysis of intelligent reflecting surface-aided decode-and-forward {UAV} communication systems,'' \emph{{IEEE} Syst. J.}, vol.~17, no.~1, pp. 246--257, Mar. 2023.

\bibitem{J:Gupta}
A.~Gupta and P.~Garg, ``Statistics of {SNR} for an indoor {VLC} system and its applications in system performance,'' \emph{{IEEE} Commun. Lett.}, vol.~22, no.~9, pp. 1898--1901, Sep. 2018.

\bibitem{J:Pan}
G.~Pan, J.~Ye, C.~Zhang, J.~An, H.~Lei, Z.~Ding, and M.~S. Alouini, ``Secure cooperative hybrid {VLC-RF} systems,'' \emph{{IEEE} Trans. Wireless Commun.}, vol.~19, no.~11, pp. 7097--7107, Nov. 2020.

\bibitem{J:Chetlur}
V.~V. Chetlur and H.~S. Dhillon, ``Downlink coverage analysis for a finite {3-D} wireless network of unmanned aerial vehicles,'' \emph{{IEEE} Trans. Commun.}, vol.~65, no.~10, pp. 4543--4558, Oct. 2017.

\bibitem{J:Nichols}
J.~Nichols and J.~Michalowicz, ``Distance distribution between nodes in a {3D} wireless network,'' \emph{J. Parallel Distrib. Comput.}, vol. 102, pp. 71--79, 2017.

\bibitem{J:Hyytia}
E.~Hyytia, P.~Lassila, and J.~Virtamo, ``Spatial node distribution of the random waypoint mobility model with applications,'' \emph{{IEEE} Trans. Mobile Comput.}, vol.~5, no.~6, pp. 680--694, Jun. 2006.

\bibitem{J:Zhong}
X.~Zhong, F.~Chen, Q.~Guan, F.~Ji, and H.~Yu, ``On the distribution of nodal distances in random wireless ad hoc network with mobile node,'' \emph{Ad Hoc Netw.}, vol.~97, p. 102026, 2020.

\bibitem{B:Mathai}
A.~M. Mathai, \emph{An Introduction to Geometrical Probability: Distributional Aspects with Applications}.\hskip 1em plus 0.5em minus 0.4em\relax Amsterdam, The Netherlands: Gordon \& Breach, 1999.

\bibitem{J:Bettstetter}
C.~Bettstetter, ``Topology properties of ad hoc networks with random waypoint mobility,'' \emph{Mob. Comput. Commun. Rev.}, vol.~7, no.~3, pp. 50--52, 2003.

\bibitem{J:Parry}
M.~Parry and E.~Fischbach, ``Probability distribution of distance in a uniform ellipsoid: Theory and applications to physics,'' \emph{J. Math. Phys.}, vol.~41, no.~4, pp. 2417--2433, 2000.

\bibitem{J:Ermolova}
N.~Y. Ermolova and O.~Tirkkonen, ``Using beta distributions for modeling distances in random finite networks,'' \emph{IEEE Commun. Lett.}, vol.~20, no.~2, pp. 308--311, Feb. 2015.

\end{thebibliography}
\end{document}